\documentclass[prl,notitlepage,aps,superscriptaddress]{revtex4-1}
\usepackage{amsmath,amsfonts,amssymb,bm}
\usepackage{epsfig}
\usepackage{graphicx}
\usepackage{wrapfig}
\usepackage{xcolor}
\usepackage{soul}

\DeclareMathOperator{\sgn}{sgn}
\begin{document}

\title{Resilient nodeless $d$-wave superconductivity in monolayer FeSe}

\author{D.F. Agterberg}
\email{agterber@uwm.edu}
\affiliation{Department of Physics, University of Wisconsin,
                 Milwaukee, WI 53201, USA}
\author{T. Shishidou}
\affiliation{Department of Physics, University of Wisconsin,
                 Milwaukee, WI 53201, USA}

\author{J. O'Halloran}
\affiliation{Department of Physics, University of Wisconsin,
                 Milwaukee, WI 53201, USA}
                 
\author{P. M. R. Brydon}
\affiliation{Department of Physics, University of Otago, P.O. Box 56,
             Dunedin 9054, New Zealand}
             
\author{M. Weinert}
\affiliation{Department of Physics, University of Wisconsin,
                 Milwaukee, WI 53201, USA}


\begin{abstract}
Monolayer FeSe exhibits the highest  transition temperature among the iron based superconductors and appears to be fully gapped, seemingly consistent with $s$-wave superconductivity. Here, we develop a theory for the superconductivity based on coupling to fluctuations of checkerboard magnetic order (which has the same translation symmetry as the lattice). The electronic states are described by a symmetry based ${\bf k}\cdot {\bf p}$-like theory and naturally account for the states observed by angle resolved photoemission spectroscopy. We show that a prediction of this theory is that the resultant superconducting state is a fully gapped, {\it nodeless}, $d$-wave state. This state, which would usually have nodes, stays nodeless because, as seen experimentally, the relevant spin-orbit coupling has an energy scale smaller than the superconducting gap.
\end{abstract}

\maketitle

The origin of superconductivity in iron based superconductors represents an important problem in condensed matter \cite{hir11,chu12}. These materials have a relatively high superconducting transition temperature ($T_c$) and reveal unconventional states which are likely a consequence of electronic interactions. The most common explanation for superconductivity originates in a repulsive interaction between electron and hole pockets, leading to a superconducting gap that changes sign between these pockets \cite{maz08,kur08}. In this context, superconductivity in single layer FeSe presents a conundrum \cite{wan12,hua17}.  Although it has the highest $T_c$ of the Fe-based superconductors, only electron pockets are present, so that the usual pairing interaction is not easily ascribable as the origin of superconductivity \cite{wan12}. Furthermore, in spite of the evidence of electronic correlations in monolayer FeSe \cite{hua17}, the observed superconducting state is consistent with a fully gapped conventional $s$-wave pairing state  \cite{fan15,zha16}.

Understanding these apparent paradoxes is complicated by the complexity of existing theoretical models of iron-based superconductors. These models contain ten orbital and two spin degrees of freedom, which often obscures the underlying physics. Here, for monolayer FeSe, we introduce a simple symmetry-based effective ${\bf k}\cdot {\bf p}$ theory containing just two orbital degrees of freedom to describe the electronic excitations at the Fermi surface. We show that when these fermions are coupled to fluctuations associated with translation invariant checkerboard magnetic (CB-AFM) order, the resultant fully gapped, nodeless, $d$-wave superconducting state naturally produces the gap anisotropy seen in angle-resolved photoemission spectroscopy (ARPES) \cite{zha16}. A key parameter  in our ${\bf k}\cdot {\bf p}$ theory is a spin-orbit coupling (SOC) energy  that is  distinct from the usual on-site SOC. This SOC would usually require the nodeless $d$-wave state to develop nodes, but
ARPES reveals this SOC is too small to allow for the nodes to develop.  We then discuss how our theory generalizes to 3D $K$-dosed bulk FeSe and to (Li$_{0.8}$Fe$_{0.2}$)OHFeSe \cite{zha16-2,du16,wen16}.
We do not include the role of interface phonons here but adhere to the viewpoint that these can enhance the $T_c$ found from other mechanisms \cite{lee14,lee15}.

In the following, we initially develop a symmetry based ${\bf k}\cdot{\bf p}$-like  theory around the ${\bf M}$-point of the Brillouin zone for the eight states (four orbital times two spin) that density functional theory (DFT) shows are relevant (we use a two-Fe unit cell throughout). We then find our first key result: when this theory is restricted to states crossing the Fermi surface, it can be understood as a simpler ${\bf k}\cdot {\bf p}$-like  theory deriving from a single four-fold degenerate spinor representation at the ${\bf M}$-point.  The band structure revealed by ARPES \cite{zha16} is consistent with this simpler theory, and we use the ARPES results to find the relevant parameters. We then develop a spin-fermion description which couples the fermions near the ${\bf M}$-point to spin fluctuations stemming from translation invariant CB-AFM order.  This leads to our second key result: this coupling naturally gives rise to a nodeless $d$-wave superconducting state with a gap anisotropy in agreement with that observed in ARPES. This result is non-trivial because nodes are expected in this $d$-wave state. Indeed, in the context of K$_x$Fe$_y$Se$_2$ superconductors,  arguments were given to show that nodes will generically appear in a related nodeless $d$-wave state \cite{maz11}.  In our case, such nodes do not appear  because the SOC  energy that would generate the nodes is observed to be too small.  Finally we discuss the extension of this theory to 3D bulk FeSe and the high $T_c$ material (Li$_{1-x}$Fe$_{x}$)OHFeSe.

\begin{figure}[htb]
\begin{center}
\includegraphics[width=0.8\textwidth]{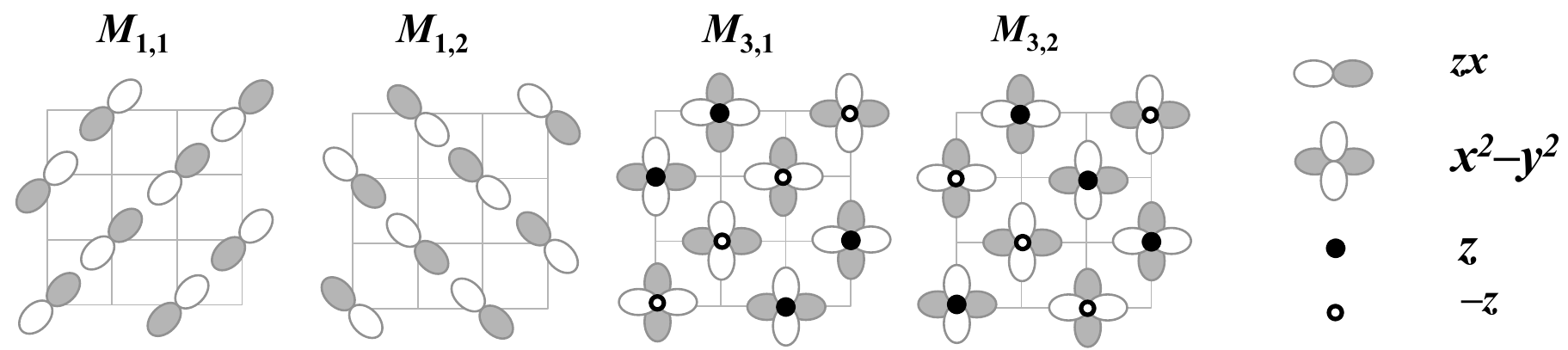}
\caption{Degenerate orbital basis sets at the $M$ point for the two electronic representations ($M_1$ and $M_3$) relevant to the electronic bandstructure near the Fermi surface. These orbitals are centered on Fe sites. Note that the Fe sites are not inversion centers, allowing for $p_z$ orbitals to mix with the $x^2-y^2$ orbitals (these $p_z$ orbitals are labeled by $\pm z$, with the sign giving the phase of the $p_z$ orbital when viewed from above).}
\end{center}
\end{figure}

{\it Effective ${\bf k}\cdot {\bf p}$ theory.}- In monolayer FeSe, the observed Fermi surfaces are close to the ${\bf M}$-point in the Brillouin zone. This motivates the development of a symmetry based ${\bf k}\cdot {\bf p}$-like theory for the electronic states near the ${\bf M}$-point. In the context of FeAs superconductors, a related theory has been developed and  we use their notation \cite{cve13} to define the relevant electronic  representations. Our DFT calculates that the states at the chemical potential are  predominantly $\{xz,yz\}$ and $x^2-y^2$ orbital states. Without SOC, the relevant linear combinations of these states that are degenerate at the ${\bf M}$-point are shown in Fig. 1.   The details of the ${\bf k}\cdot {\bf p}$ theory for these four orbitals is given in the Supplemental Material \cite{SM}. We do not display the full ${\bf k}\cdot {\bf p}$ theory here since a key simplification arises when we restrict this more general ${\bf k}\cdot {\bf p}$ theory to the bands that cross the chemical potential: we find that the resultant simplified ${\bf k}\cdot {\bf p}$ theory  has the same structure {\it as if we had kept only the $x^2-y^2$ $M_3$ orbital basis}. Note that this does not imply that these states are pure $x^2-y^2$ orbitals, indeed we find that in general there is a linear combination of $x^2-y^2$, $xz$ and $yz$ orbitals.  The reason the symmetry of the $x^2-y^2$ orbitals dictate the effective ${\bf k}\cdot {\bf p}$ theory stems from the following three observations: i) along the $\Gamma$ to ${\bf M}$ direction ($k_x=k_y$), DFT shows that the $M_{1,1}$ and $M_{3,2}$ states cross the Fermi surface; and ii) when $k_x=k_y\ne 0$, the $M_{1,1}$ and the $M_{3,1}$ states belong to the same irreducible representation along the direction from $\Gamma$ to ${\bf M}$ (so that states crossing the Fermi surface have the same symmetry as $M_{3,1}$ and $M_{3,2}$). We note this effective theory also applies when there are strong orbital renormalizations, such as in Ref.~\onlinecite{yi15}.

To construct the effective ${\bf k}\cdot {\bf p}$ theory it is useful to note that the bilinear products of the electronic operators that appear in the Hamiltonian can  be assigned to irreducible representations at the $\Gamma$-point. To this end, we introduce $\Psi_{\bf k}$ as a four component spinor with two orbital degrees of freedom described by $\tau_i$ Pauli matrices (these effective $x^2-y^2$ orbitals are  ${\bf k}$-dependent linear combinations of the $\{xz,yz\}$ and $x^2-y^2$ orbitals shown in Fig.1) and two spin degrees of freedom described by $\sigma_i$ Pauli matrices. To construct the Hamiltonian, we identify the symmetries of the $\tau_i$ and $\sigma_i$ operators together with the symmetries of $k$-dependent functions in Table 1. 

\begin{table}
\begin{tabular}{|c|c|c|c|c|c|c|c|c|}
    \hline
   $P_{\Gamma}$&$f({\bf k})$&$\tau_j$&$\sigma_i$&$P_{\Gamma}$&$f({\bf k})$&$\tau_j$\\
   \hline
   $A_{1g}$ &  $c, k_x^2+k_y^2$ &$\tau_0$ & - &$A_{1u}$&-&$\tau_y$\\
   $A_{2g}$  & - &- &  $\sigma_z$ &$A_{2u}$&-&-\\
   $B_{1g}$  & $k_x^2-k_y^2$&-  & - &$B_{1u}$&-&-\\
   $B_{2g}$  & $k_xk_y$ &$\tau_z$ &  - &$B_{2u}$&-&$\tau_x$\\
   $E_{g}$  & - & - & $\{\sigma_{x},\sigma_y\}$&$E_u$&$\{k_x,k_y\}$&- \\
\hline
\end{tabular}
\caption{Symmetry of functions $f({\bf k})$ and the operators $\tau_j$ and $\sigma_i$ used in the ${\bf k}\cdot {\bf p}$-like single particle Hamiltonian. These are labeled by $P_{\Gamma}$, characterizing irreducible representations at the $\Gamma$-point.}
\label{table1}
\end{table}

\begin{equation}
H_M=\sum_{\bf k}\Psi^{\dagger}_{\bf k}\Bigg\{ \epsilon_0({\bf k}) +\gamma_{xy}({\bf k})\tau_z+\tau_x\big[\gamma_y({\bf k})\sigma_x+\gamma_x({\bf k})\sigma_y\big]\Bigg\}\Psi_{\bf k},
\label{hkp}
\end{equation}
here $\gamma_{xy}({\bf k})$ has $B_{2g}$ symmetry ($k_xk_y$-like), and $\{\gamma_x({\bf k}),\gamma_y({\bf k})\}$ have $E_u$ symmetry ($\{k_x,k_y\}$-like). It is important to note that symmetry dictates no other matrices of the form $\tau_i\sigma_j$ can appear in Eq.~\ref{hkp}.  As shown below, this ensures that only the $\tau_x$ term is key to understanding the nodeless $d$-wave state. This term originates from SOC, but is not the usual on-site SOC that is often discussed  \cite{cve13,bor16}.

 An examination of the observed band structures shows that the bands can be understood as deriving from a {\it single} {\bf M}-point spinor representation \cite{bor16,yi15}, indicating that  Eq.~\ref{hkp} can be extended to the lower energy states at the ${\bf M}$-point. Choosing $\epsilon_0({\bf k})=\frac{k_x^2+k_y^2}{2m}-\mu$, $\gamma_{xy}({\bf k})=a k_xk_y$ and
 $\{\gamma_x({\bf k}),\gamma_y({\bf k})\}=\{v_{so}k_x,v_{so}k_y\}$ in Eq.~\ref{hkp} reproduces the bands and the Fermi surface observed by ARPES \cite{zha16} when the parameters are chosen as $\mu=55$ meV,  $1/2m=1375$ meV {\AA}$^2$, $a =600$ meV {\AA}$^2$, and $|v_{so}|\le 15$ meV {\AA}.  Importantly, spin-orbit splitting was not observed in \cite{zha16}; the 5meV energy resolution of this experiment thus provides an upper bound on $v_{so}$, but within these limits we shall treat it as a variable. In the following we will use this parameterization which leads to a typical Fermi wavevector $k_0=0.2$ {\AA}$^{-1}$.

{\it Spin fluctuations.}- Next we couple these ${\bf M}$-point fermions to spin fluctuations. Our approach is to follow a spin-fermion model \cite{aba03}, with the coupling determined by symmetry arguments.  Fermions near the $\Gamma$ point are not included here, this is justified since they are 80 meV below the Fermi energy \cite{hua17,fan15,zha16} (see, however Refs.~\cite{hua17,che15,chu16}),  this implies that the usual stripe antiferromagentic fluctuations \cite{hir11,chu12} cannot play a role in superconductivity since they couple ${\bf M}$ point and $\Gamma$ point fermions. To couple states on the ${\bf M}$-point Fermi surfaces, magnetic fluctuations must have small ${\bf q}$ (on the order of $k_0$). This suggests that the relevant magnetic order does not break the translation symmetry of the lattice. DFT reveals that the only realistic possibility is fluctuations associated with CB-AFM order \cite{wan16,coh15,FeSTO16,shi17}. This is consistent with experiment \cite{wan16-2, ma17}. In particular, in bulk 3D FeSe (with a low $T_c=8$ K), CB-AFM and stripe magnetic fluctuations are both observed, and the onset of nematic order suppresses the CB-AFM fluctuations \cite{wan16-2}. Furthermore, in (Li$_{0.8}$Fe$_{0.2}$)ODFeSe ($T_c=39$ K), which has no nematic order, the lowest energy spin excitations are consistent with nesting of the ${\bf M}$-point Fermi surfaces \cite{ma17}. Due to the two-iron unit cell, CB-AFM order is translation invariant. In the ordered state, the moments are opposite on the two Fe atoms. This implies that CB-AFM order breaks time-reversal and  inversion symmetries (the inversion center lies between the two Fe sites). The ordered state has a spatial $B_{2u}$ symmetry and breaks spin-rotational invariance. Including all coupling terms allowed by symmetry (see the Supplemental Material \cite{SM}) and projecting onto the states near the chemical potential yields the following coupling \begin{equation}
g\sum_{{\bf k},{\bf q}}f({\bf k}){\bf S}_{-{\bf q}}\cdot  \Psi^{\dagger}_{{\bf k}+{\bf q}/2}\tau_x\vec{\sigma}\Psi_{{\bf k}-{\bf q}/2}
\label{coupling}
\end{equation}
where, for ${\bf k}$ near the ${\bf M}$-point, $f({\bf k})=1+\alpha(k_x^2+k_y^2)/k_0^2$. Note that when no SOC is present, magnetic fluctuations only couple fermions on different bands. This is apparent from Eq.~\ref{hkp}, for which the bands are eigenstates of $\tau_z$ when $\gamma_x=\gamma_y=0$, and from Eq.~\ref{coupling} which only couples different eigenstates of $\tau_z$. This property leads to the nodeless $d$-wave superconducting state. To complete the theory, we need to include a  susceptibility for the magnetic fluctuations. We assume that this takes the static form
 \begin{equation}
 \sum_{\bf q}\chi_0^{-1}({\bf q}){\bf S}_{{\bf q}}\cdot {\bf S}_{-{\bf q}}
 \label{susc}
 \end{equation}
 with $\chi_0({\bf q})=\chi_0/(\xi^{-2}+q^2)$.
Without SOC, Eqs.~\ref{coupling} and \ref{susc} will lead to a dynamically generated Landau-damping-like term in the spin susceptibility similar to  spin-fermion theories with hot spots \cite{aba03,met10,sci17,ber12}. When SOC is non-zero, these hot spots become hot Fermi surfaces and the spin susceptibility will develop a true Landau damping \cite{del07,sch16}. Our goal is to understand what superconducting pairing these fluctuations give rise to. We therefore use a weak-coupling limit, for which the spin dynamics are not important \cite{aba03}. It will nevertheless be interesting to examine this theory in the stronger coupling regime. We set $q^2=0$ in the static susceptibility (this will not qualitatively change the results). The resultant theory leads to a robust prediction for the pairing symmetry and gap anisotropy.

{\it Superconductivity.}- In the weak-coupling limit, the effective electron-electron interaction becomes
\begin{equation}
-g^2\chi_0\xi^2\sum_{{\bf k},{\bf k}',{\bf q}} f({\bf k})f({\bf k}')\Psi^{\dagger}_{{\bf k}+{\bf q}/2}\tau_x \vec{\sigma} \Psi_{{\bf k}-{\bf q}/2}\cdot \Psi^{\dagger}_{{\bf k}'-{\bf q}/2}\tau_x \vec{\sigma} \Psi_{{\bf k}'+{\bf q}/2}.
\end{equation}
Initially,  we consider no SOC. In this case, the Fermi surface consists of two co-centered ellipses with the second ellipse found by rotating the first by $\pi/2$. We define the gaps $\Delta_{\pm}({\bf k})$ on these two ellipses. Assuming usual intra-band Cooper pairs and spin-singlet pairing (note that with non-zero SOC, the spin-singlet pairing considered here will generally mix with an even parity spin-triplet pairing \cite{cve13}), we find the linear gap equation
\begin{equation}
\Delta_{\pm}({\bf k})=-VT\sum_{{\bf k}',\omega_n}\frac{f^2(\frac{{\bf k}+{\bf k}'}{2})\Delta_{\mp}({\bf k}')}{\epsilon_{\mp}^2({\bf k}')+\omega_n^2}
\end{equation}
where the effective  interaction $V$ is repulsive and $\epsilon_{\pm}({\bf k})=\epsilon_0({\bf k})\pm \gamma_{xy}({\bf k})$. To solve the gap equation, it is useful to rescale the elliptical bands so that the constant energy surfaces become circles. This is done by setting $\tilde{k}_x=k_x/\epsilon$ and $\tilde{k}_y=\epsilon k_y$ on the first band and $\tilde{k}_x=k_x \epsilon$ and $\tilde{k}_y=k_y/\epsilon$ on the second band (with $\epsilon=1.13$). For ${\bf k}$ on the Fermi surface, the functions $\Delta_{\pm}({\bf k})$ become $\Delta_{\pm}(\phi)$ where $\phi$ is the angle with respect to the $k_x$ axis. We find the solution with the highest transition temperature satisfies
\begin{equation}
\left(
  \begin{array}{cc}
   \Delta_+ (\phi)& 0 \\
    0 &  \Delta_- (\phi)\\
  \end{array}
\right)=\Delta_d(\phi)\tau_0+\Delta_z(\phi)\tau_z
\end{equation}
where $\Delta_d(\phi)=\Delta_2 \sin 2\phi$ and $\Delta_z=\Delta_0+\Delta_4 \cos 4\phi$. More specifically, for $f({\bf k})=1+0.6(k_x^2+k_y^2)/k_0^2$, we find $\Delta_{2}/\Delta_0=-0.11$ which agrees with experiment ($\Delta_{2}/\Delta_0=-0.12$ \cite{zha16}), and $\Delta_{4}/\Delta_0=0.0015$, which is two orders of magnitude smaller than experiment.  We show later that a larger $\Delta_{4}$ can be generated by the SOC. The gap structure and Fermi surface are plotted in Fig.2. We note that this is a nodeless $d_{xy}$ gap (this is a nodeless $d_{x^2-y^2}$ gap in a one-Fe unit cell), for which nodes would usually be expected along the lines $k_x=0$ and $k_y=0$.
\begin{figure}[htb]
\begin{center}
\includegraphics[width=0.4\columnwidth]{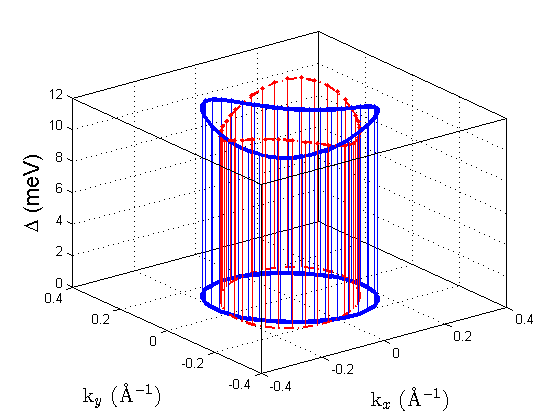}	
\caption{Gap anisotropy and Fermi surface found without SOC. The gaps on the two ellipses are of opposite sign.}
\end{center}
\end{figure}

To examine the role of SOC, we consider the Bogoliubov deGennes (BdG) equations associated with the gap structure discussed above (see also Ref.~\onlinecite{agt17})
\begin{equation}
H_{BdG}=\left(
          \begin{array}{cc}
            \epsilon_0\tau_0\sigma_0+\gamma_{xy}\tau_z\sigma_0+\tau_x(\gamma_y\sigma_x+\gamma_x\sigma_y) & (\Delta_d\tau_0+\Delta_z\tau_z)i\sigma_y \\
           -(\Delta_d\tau_0+\Delta_z\tau_z)i\sigma_y & -\epsilon_0\tau_0\sigma_0-\gamma_{xy}\tau_z\sigma_0+\tau_x(\gamma_y\sigma_x-\gamma_x\sigma_y) \\
          \end{array}
        \right)
        \label{Hbdg}
\end{equation}
where $\epsilon_0$, $\gamma_{xy}$, $\gamma_x$, and $\gamma_y$ are as defined above and we take $\Delta_d=\Delta_2 k_xk_y/k_0^2$ and $\Delta_z=\Delta_0$ with the values $\Delta_0=11$ meV and $\Delta_2=-1.5$ meV to compare to experiment. The exact quasiparticle dispersion can be found for Eq.~\ref{Hbdg},
\begin{equation}
E_{\pm}(k)=\sqrt{\epsilon_0^2+\gamma_{xy}^2+\gamma_x^2+\gamma_y^2+\Delta_d^2+\Delta_z^2\pm2\sqrt{(\epsilon_0\gamma_{xy}+\Delta_d\Delta_z)^2+(\gamma_x^2+\gamma_y^2)(\epsilon_0^2+\Delta_z^2)}}.
\label{qp}
\end{equation}
Prior to examining Eq.~\ref{qp}, it is useful to numerically examine the case of strong SOC, $v_{so}=80$ meV {\AA}, which leads to $v_{so}k_0=16$ meV, which is larger than the superconducting gap. The resultant Fermi surface and gap anisotropy are shown in Fig.~3. In this case, nodes develop along the $k_x=0$ and $k_y=0$ directions. However, if we take a much weaker SOC, $v_{so}=12$ meV {\AA} (corresponding to $v_{so}k_0=2.4$ meV which is consistent with experiment \cite{zha16}), then we get the Fermi surface and gap anisotropy shown in Fig~4. Now the nodes have been removed, and replaced by local gap minima, much like what is  seen in ARPES measurements \cite{zha16}.

\begin{figure}[htb]
\begin{center}
\includegraphics[width=0.45\columnwidth]{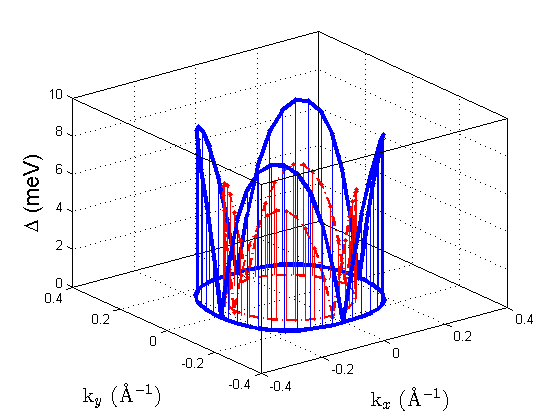}	
\end{center}
\caption{Gap anisotropy and Fermi surface for $v_{so}=80$ meV {\AA}. Even though the nodes appear to sit on the Fermi surfaces, they are actually located between them.}
\end{figure}

\begin{figure}[htb]
\begin{center}
\includegraphics[width=0.45\columnwidth]{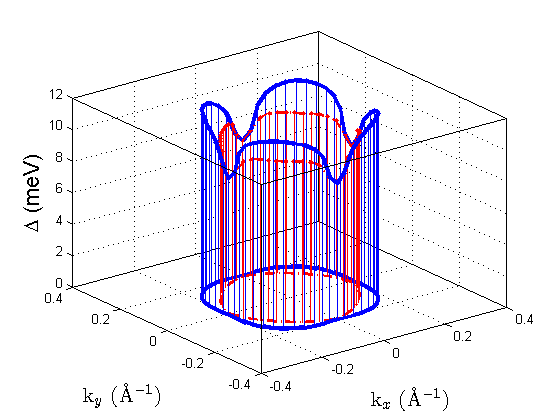}	
\end{center}
\caption{Gap anisotropy and Fermi surface for $v_{so}=12$ meV {\AA}. Plotted is the minimum gap value, which lies near the Fermi surface, but not on it. Along the directions $k_x=0$ and $k_y=0$, the minimum gap lies midway between the Fermi surfaces.}
\end{figure}

To understand when nodes appear, we examine Eq.~\ref{qp} along the nodal direction $k_x=0$. In this case, $\gamma_{xy}=\Delta_d=\gamma_x=0$, and Eq.~\ref{qp} yields $E_{\pm}=|\sqrt{\epsilon_0^2+\Delta_z^2}\pm \gamma_y|$, rewriting $\epsilon_0=\epsilon_{\pm}\mp|\gamma_y|$ and solving $E_{-}=0$ for $\epsilon_{\pm}$, yields the two nodal conditions $\epsilon_{\pm}=\pm(|\gamma_y|-\sqrt{\gamma_y^2-\Delta_z^2})$. Consequently, once the SOC $|\gamma_y|<\Delta_0$, the nodes disappear. Given that ARPES \cite{zha16} reveals the SOC energy is smaller than the gap, a nodeless $d_{xy}$ pairing state is expected. A nodeless $d$-wave state has also been discussed in the context of the cuprates \cite{zhu16} and also in the context of Fe-based superconductors \cite{nic17,li16,chu16-2,kan16}. In both these cases, the relevant coupling that removes the nodes was not a SOC and hence not necessarily small. In our case, the experimentally observed smallness of the relevant SOC naturally gives rise to a nodeless $d$-wave state. The nodeless $d$-wave state is possible due to existence of two bands. The nodes of each of these two bands can annihilate, which is not possible for a single-band superconductor. Note that this analysis implies the gap minima depend strongly upon the ratio of $v_{so} k_0/\Delta_0$. If it is possible to vary this ratio, perhaps through electric and magnetic fields, pressure, or temperature, the gap minima should vary relative to the gap maxima. In the Supplemental Material \cite{SM}, we have included the role of symmetry breaking by the interface. 


 {\it 3D materials FeSe and (Li$_{1-x}$Fe$_{x}$)OHFeSe.-} It has been argued that the high $T_c$'s in electron doped  K-dosed 3D FeSe  and in (Li$_{1-x}$Fe$_{x}$)OHFeSe have an origin similar to that in monolayer FeSe \cite{zha16-2,du16,wen16}. It is therefore reasonable to understand how the above analysis differs for these materials. The key difference is in 3D a $c$-axis dispersion exists. This can modify the single particle Hamiltonian through the inclusion of $\cos (c k_z/\pi)$ and $\sin (c k_z/\pi)$ where $c$ is the $c$-axis lattice spacing. Symmetry arguments reveal that $\sin (c k_z/\pi)$ cannot appear and that $\cos(c k_z/\pi)$ is allowed to modify all existing parameters ($\epsilon_0, \gamma_{xy},\gamma_x$, and $\gamma_y$).  A second difference is the appearance of nematic order in 3D bulk FeSe.  Symmetry reveals that this allows three additional terms in the Hamiltonian: $k_xk_y\tau_0\sigma_0$, $\tau_z\sigma_0$, and $\tau_x(k_x\sigma_x+k_y\sigma_y)$. Importantly, these new terms are also equivalent to modifying the parameters $\epsilon_0, \gamma_{xy},\gamma_x$, and  $\gamma_y$. Hence Eq. 10, with modified $\epsilon_0, \gamma_{xy},\gamma_x$, and $\gamma_y$ will still yield the exact quasiparticle spectrum, so our qualitative results are unchanged. That is, if the SOC energy is smaller than $\Delta$ for all $k_z$, a nodeless gap will result.  Note that the relevant SOC is again that given by $\gamma_i({\bf k})$ in Eq.~\ref{hkp}, which can be substantially  smaller than the usual on-site SOC (estimated to be 25 meV in 3D FeSe \cite{bor16}). We expect similar considerations will apply to K$_x$Fe$_2$Se$_2$ \cite{guo10,maz11} and Li$_x$(NH$_3$)$_y$Fe$_2$Se$_2$ \cite{sun17} however, the different space group(I4/mmm) with respect to K-dosed 3D  FeSe and (Li$_{1-x}$Fe$_{x}$)OHFeSe (P4/nmm) necessitates a more detailed analysis.

{\it Conclusions.}- We have developed a simple and realistic spin-fermion model to describe high-$T_c$ superconductivity in monolayer FeSe on SrTiO$_3$. This model accounts for the observed band structure and also naturally gives a gap structure that agrees with experiment.  The predicted state is a nodeless $d$-wave superconductor. The expected nodes for this state develop only if the relevant SOC energy is larger than the superconducting gap, which is experimentally observed not to be the case.  A careful experimental examination of the evolution of the gap minima in response to external fields, temperature, or pressure can be used to verify this nodeless $d$-wave superconducting state.


{\it Acknowledgements.}-  D.F.A., T.S., J.O., and M.W were  supported  by  the
National  Science  Foundation  Grant  No.  DMREF-
1335215. DFA was also funded  by the Gordon and Betty Moore Foundation’s EPiQS Initiative through Grant GBMF4302. We thank Tao Jia, Steve Kivelson, Lian Li, Igor Mazin, Rob Moore, Sri Raghu, Slavko Rebec, ZX Shen, and Carsten Timm for useful discussions.

\onecolumngrid


\clearpage
\onecolumngrid

\setcounter{equation}{0}
\setcounter{page}{1}

\begin{center}
\textbf{\large  Supplemental Material for\\[0.5ex]
Resilient  nodeless $d$-wave superconductivity in monolayer FeSe}\\[2ex]
D.F. Agterberg, T. Shishidou, J. O'Halloran, P.M.R. Brydon, 
M. Weinert
\end{center}

\section{Four band ${\bf kp}$ theory}

We label the two-dimensional $\{xz,yz\}$ and $x^2-y^2$ representations as $M_1$ and $M_3$. Symmetry arguments can be used to construct the ${\bf k}\cdot{\bf p}$-like  theory for these states. A key simplification  follows from the observation that the Hamiltonian depends upon bilinear products of the electronic operators and that these bilinear products can  be assigned to irreducible representations at the $\Gamma$-point. We use three sets of Pauli matrices to define these operators: $\Gamma_i$  matrices describe the two representational degrees of freedom ($M_1$,$M_3$), the $\tilde{\tau}_i$ matrices describe the two orbital degrees of freedom within the representations, and the $\sigma_i$  matrices describe the two spin degrees of freedom. In Table 1, using the definition of the $\Gamma$-point representations as defined in \cite{cve13}, we give the corresponding symmetries of the operators that define the single-particle Hamiltonian.

Without SOC, the ${\bf k}\cdot {\bf p}$ like Hamiltonian is the same as found as in Ref.~\onlinecite{cve13}, and we write it as
\begin{equation}
H_{CV}=\sum_{\bf k} \tilde{\Psi}^{\dagger}_{\bf k}\Bigg[\frac{\Gamma_0+\Gamma_z}{2}(\epsilon_1\tilde{\tau}_0+a_1k_xk_y\tilde{\tau}_z)+\frac{\Gamma_0-\Gamma_z}{2}(\epsilon_3\tilde{\tau}_0+a_3k_xk_y\tilde{\tau}_z)
+v\Gamma_y(k_x\tilde{\tau}_0+k_y\tilde{\tau}_z)\Bigg]\tilde{\Psi}_{\bf k}
 \label{CV}
\end{equation}
with $\epsilon_{1,3}({\bf k})=c_{1,3}+(k_x^2+k_y^2)/(2m_{1,3})-\mu$, $\tilde{\Psi}_{{\bf k}}$ is an eight-component spinor. We extend this to include SOC,
\begin{equation}
H_{so}=\sum_{\bf k} \tilde{\Psi}^{\dagger}_{\bf k}\Bigg[\lambda_1\frac{\Gamma_0+\Gamma_z}{2}\tilde{\tau}_x(k_y\sigma_x-k_x\sigma_y)+\lambda_2\frac{\Gamma_0-\Gamma_z}{2}\tilde{\tau}_x\big(k_y\sigma_x+k_x\sigma_y\big)+
\lambda(\Gamma_y\tilde{\tau}_x\sigma_y+\Gamma_x\tilde{\tau}_y\sigma_x)\Bigg]\tilde{\Psi}_{\bf k}.
\end{equation}
The last term in  $H_{so}$ has also been found in \cite{cve13}. Note that once SOC is included, the relevant orbital states are mixed for all ${\bf k}$,  even at the ${\bf M}$-point. Formally, this implies that there is only a single 4-fold degenerate irreducible double group representation at the ${\bf M}$-point (as opposed to four such representations without SOC).


Now we proceed to develop a description of the states at the Fermi surface by assuming that the energy scales of $H_{so}$ are smaller than those of $H_{CV}$. We therefore diagonalize $H_{CV}$ and project $H_{so}$ onto the two bands that cross the Fermi surface. Ensuring that the eigenstates at the chemical potential are chosen to be continuous with ${\bf k}$  yields the effective Hamiltonian for these two bands
\begin{equation}
H_M=\sum_{\bf k}\Psi^{\dagger}_{\bf k}\Bigg\{ \epsilon_0({\bf k}) +\gamma_{xy}({\bf k})\tau_z+\tau_x\big[\gamma_y({\bf k})\sigma_x+\gamma_x({\bf k})\sigma_y\big]\Bigg\}\Psi_{\bf k},
\label{hkp}
\end{equation}
where

\begin{equation}
\epsilon_0=\frac{\epsilon_1+\epsilon_3}{2}+ \frac{E_+ + E_-}{2}, 
\end{equation}
\begin{equation}
\gamma_{xy}=\frac{a_1+a_3}{2} k_x k_y + \frac{E_+ - E_-}{2}, 
\end{equation}

\begin{eqnarray}
\gamma_x=&&\lambda_1\frac{k_x\sgn(k_y^2-k_x^2)}{2}\sqrt{\Big(1+\frac{\gamma_+}
{E_+}\Big)\Big(1+\frac{\gamma_-}{E_-}\Big)}
+
\lambda_2\frac{k_x}{2}\sqrt{\Big(1-\frac{\gamma_+}
{E_+}\Big)\Big(1-\frac{\gamma_-}{E_-}\Big)} \nonumber \\&&+\lambda
 \sgn(v)
\Big[\frac{\sgn(k_x+k_y)
}{2}\sqrt{\Big(1+\frac{\gamma_+}
{E_+}\Big)\Big(1-\frac{\gamma_-}{E_-}\Big)}+\frac{\sgn(k_x-k_y)
}{2}\sqrt{\Big(1-\frac{\gamma_+}
{E_+}\Big)\Big(1+\frac{\gamma_-}{E_-}\Big)}\Big]
\end{eqnarray}

\begin{eqnarray}
\gamma_y=&&\lambda_1\frac{k_y\sgn(k_x^2-k_y^2)}{2}\sqrt{\Big(1+\frac{\gamma_+}
{E_+}\Big)\Big(1+\frac{\gamma_-}{E_-}\Big)}
+
\lambda_2\frac{k_y
}{2}\sqrt{\Big(1-\frac{\gamma_+}
{E_+}\Big)\Big(1-\frac{\gamma_-}{E_-}\Big)} \nonumber \\&&+\lambda
\ \sgn(v)
\Big[\frac{\sgn(k_x+k_y)
}{2}\sqrt{\Big(1+\frac{\gamma_+}
{E_+}\Big)\Big(1-\frac{\gamma_-}{E_-}\Big)}-\frac{\sgn(k_x-k_y)
}{2}\sqrt{\Big(1-\frac{\gamma_+}
{E_+}\Big)\Big(1+\frac{\gamma_-}{E_-}\Big)}\Big]
\end{eqnarray}
where $\gamma_{\pm}=(\epsilon_1-\epsilon_3\pm(a_1-a_3)k_xk_y)/2$ and $E_{\pm}=\sqrt{\gamma_{\pm}^2+v^2(k_x\pm k_y)^2}$.  

Basis functions of $\Psi_{\bf k}$ are
\begin{equation}
A_+({\bf k}) | M_{1,1}\rangle + B_+({\bf k}) | M_{3,1}\rangle
\hspace{2mm}  \textrm{and} \hspace{2mm}
A_-({\bf k}) | M_{1,2}\rangle + B_-({\bf k}) | M_{3,2}\rangle
\end{equation}
with 
\begin{eqnarray}
A_\pm ({\bf k}) &=& - i \ \sgn(v)\sgn(k_x \pm k_y ) \sqrt{\frac{1}{2}\left(1+\frac{\gamma_\pm}{E_\pm}\right)},\\
B_\pm ({\bf k}) &=&  \sqrt{\frac{1}{2}\left(1-\frac{\gamma_\pm}{E_\pm}\right)}. 
\end{eqnarray}

Some algebra reveals that $\gamma_{xy}({\bf k})$ has $B_{2g}$ symmetry ($k_xk_y$-like) and $\{\gamma_x({\bf k}),\gamma_y({\bf k})\}$ have $E_u$ symmetry (
$\{k_x,k_y\}$-like).  Notice that there is no on-site SOC in Eq.~\ref{hkp}. This is because the on-site SOC mixes two different ${\bf M}$-point representations and we have only kept 
the single ${\bf M}$-point representation that is relevant near the chemical potential (if an on-site SOC did exist for a single ${\bf M}$-point representation, this would split the 4-fold degeneracy that is required by symmetry).
Instead of using the detailed expressions for the parameters and fitting to our DFT results, we fit the coefficients to ARPES data \cite{zha16}.  In particular, we set 
$\epsilon_0({\bf k})=\frac{k_x^2+k_y^2}{2m}-\mu$, $\gamma_{xy}({\bf k})=a k_xk_y$ and $\{\gamma_x({\bf k}),\gamma_y({\bf k})\}=\{v_{so}k_x,v_{so}k_y\}$.  Eq.~\ref{hkp} matches the bands and the Fermi surface observed by ARPES when the parameters are chosen as $\mu=55$ meV,  $1/2m=1375$ meV {\AA}$^2$, $a =600$ meV {\AA}$^2$, and $|v_{so}|\le 15$ meV {\AA}. Choosing  $v_{so}=12$ meV {\AA} yields the band structure shown in Fig.1 and the Fermi surface shown in Fig.2.

\begin{figure}[htb]
\begin{center}
\includegraphics[width=0.4\columnwidth]{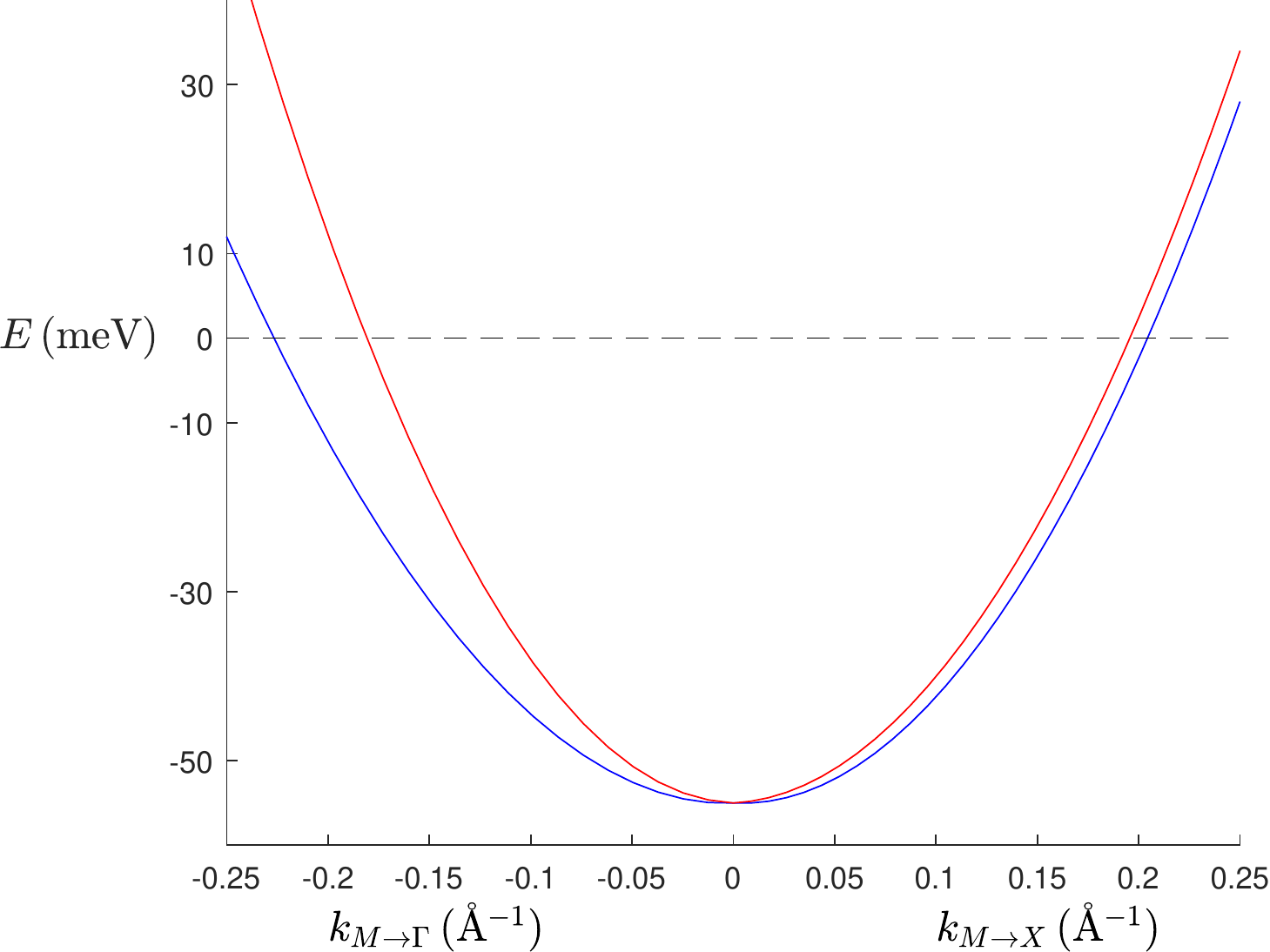}	
\caption{Bands along high-symmetry directions for the effective ${\bf k}\cdot {\bf p}$ theory. Here $v_{so}=12$ meV {\AA}.}
\end{center}
\end{figure}

\begin{figure}[htb]
\begin{center}
\includegraphics[width=0.4\columnwidth]{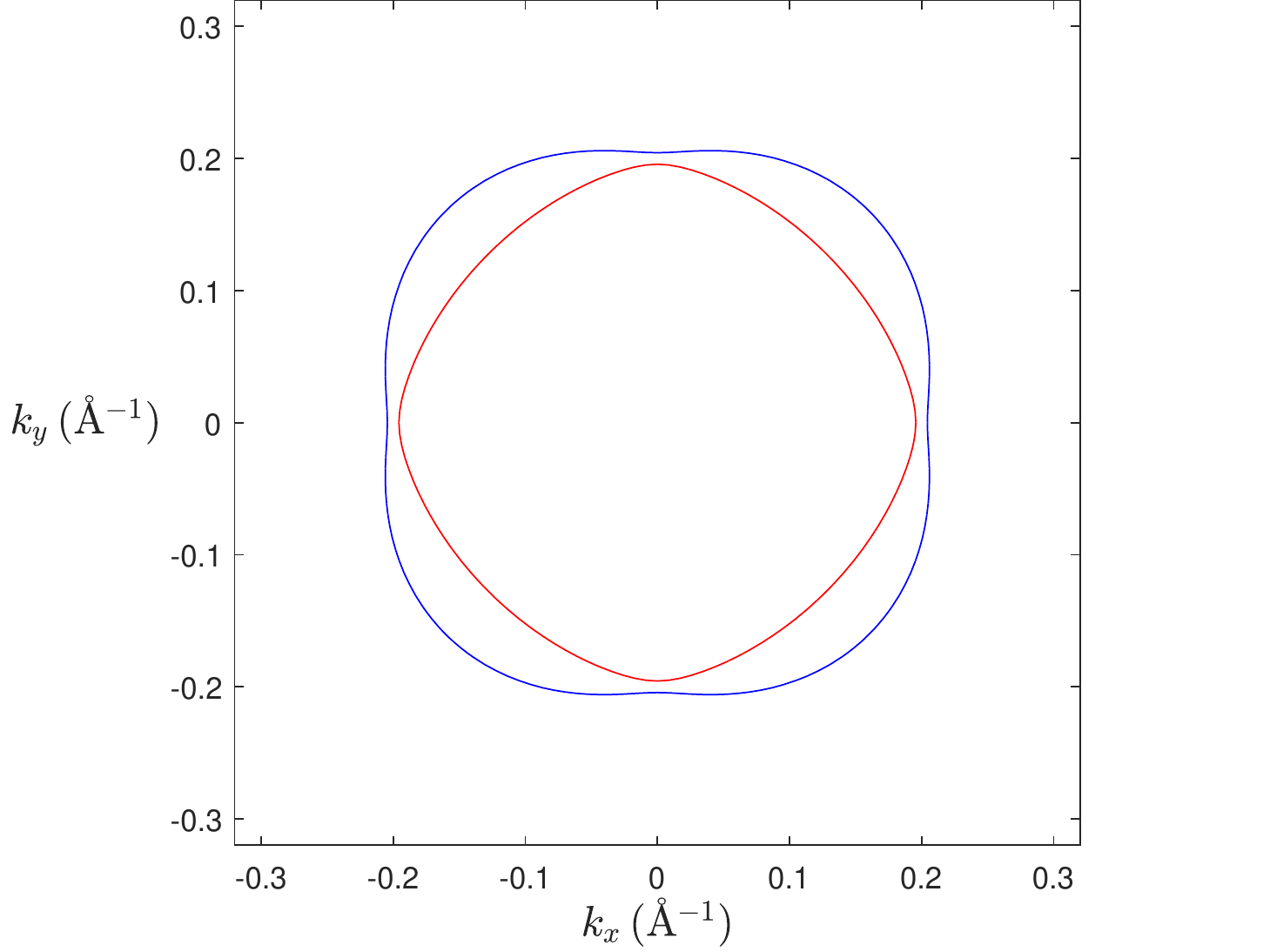}	
\caption{Fermi surface for the effective ${\bf k}\cdot {\bf p}$ theory. Here $v_{so}=12$ meV {\AA}.}
\end{center}
\end{figure}

\section{Coupling to Spin Fluctuations}

Prior to projecting to the states near the chemical potential, the full symmetry allowed coupling to spin-fluctuations is 
\begin{equation}
\sum_{{\bf k},{\bf q}}{\bf S}_{-{\bf q}}\cdot  \tilde{\Psi}^{\dagger}_{{\bf k}+{\bf q}/2}\Bigg[g_1\frac{\Gamma_0+\Gamma_z}{2}\tilde{\tau}_x\frac{(k_x^2-k_y^2)}{k_0^2}+g_2\frac{\Gamma_0-\Gamma_z}{2}\tilde{\tau}_x+
g_3\frac{k_x\Gamma_y\tilde{\tau}_x+k_y\Gamma_x\tilde{\tau}_y}{k_0}\Bigg]\vec{\sigma}\tilde{\Psi}_{{\bf k}-{\bf q}/2}.
\end{equation}
where $k_0=0.2$ {\AA}$^{-1}$ is approximately the Fermi wavevector.

\begin{table}
\begin{tabular}{|c|c|c|c|c|c|c|c|c|}
   \hline
   $P_{\Gamma}$&$f({\bf k})$&$\Gamma_i\tilde{\tau}_j$&$\sigma_i$&$P_{\Gamma}$&$f({\bf k})$&$\Gamma_i\tilde{\tau}_j$\\
   \hline
   $A_{1g}$ &  $c, k_x^2+k_y^2$ &$\tilde{\tau}_0(\Gamma_0\pm \Gamma_z)$ & - &$A_{1u}$&-&$(\Gamma_0-\Gamma_z)\tilde{\tau}_y$\\
   $A_{2g}$  & - &- &  $\sigma_z$ &$A_{2u}$&-&$(\Gamma_0+\Gamma_z)\tilde{\tau}_x$\\
   $B_{1g}$  & $k_x^2-k_y^2$&-  & - &$B_{1u}$&-&$(\Gamma_0+\Gamma_z)\tilde{\tau}_y$ \\
   $B_{2g}$  & $k_xk_y$ &$(\Gamma_0\pm \Gamma_z)\tilde{\tau}_z$ &  - &$B_{2u}$&-&$(\Gamma_0-\Gamma_z)\tilde{\tau}_x$\\
   $E_{g}$  & - & $\{\Gamma_x\tilde{\tau}_y,\Gamma_y\tilde{\tau}_x\}$,$\{-\Gamma_y\tilde{\tau}_y,\Gamma_x\tilde{\tau}_x\}$& $\{\sigma_{x},\sigma_y\}$&$E_u$&$\{k_x,k_y\}$&$\{\Gamma_y\tilde{\tau}_0,\Gamma_y\tilde{\tau}_z\}$,$\{\Gamma_x\tilde{\tau}_0,\Gamma_x\tilde{\tau}_z\}$ \\
\hline
\end{tabular}
\caption{Symmetry of functions $f({\bf k})$ and the operators $\Gamma_i\tilde{\tau}_j$ and $\sigma_i$ used in the ${\bf k}\cdot {\bf p}$-like single particle Hamiltonian. These are labeled by $P_{\Gamma}$, characterizing irreducible representations at the $\Gamma$-point.}
\label{table1}
\end{table}

\section{Interface Symmetry Breaking}

To illustrate the utility of our effective theory, we include the effects of the interface on superconductivity. In particular, the interface removes the mirror glide plane symmetry and formally allows terms with  $A_{2u}$ symmetry in the Hamiltonian.  This leads to an additional term in the ${\bf k}\cdot {\bf p}$ theory; $M_I\tau_x$ with $M_I=\lambda_I (k_x^2-k_y^2$). The resultant Bogoliubov deGennes (BdG) equation is now 
\begin{equation}
H_{BdG}=\left(
          \begin{array}{cc}
            \epsilon_0\tau_0\sigma_0+\gamma_{xy}\tau_z\sigma_0+\tau_x(M_I+\gamma_y\sigma_x+\gamma_x\sigma_y) & (\Delta_d\tau_0+\Delta_z\tau_z)i\sigma_y \\
           -(\Delta_d\tau_0+\Delta_z\tau_z)i\sigma_y & -\epsilon_0\tau_0\sigma_0-\gamma_{xy}\tau_z\sigma_0+\tau_x(M_I+\gamma_y\sigma_x-\gamma_x\sigma_y) \\
          \end{array}
        \right)
        \label{Hbdg}
\end{equation}
The normal state Fermi surface in this case consists of four separate Fermi surfaces. Adding the interface coupling still allows for an exact solution of the quasi-particle spectrum in the superconducting state 
\begin{equation}
E_{\pm,\pm}(k)=\sqrt{\epsilon_0^2+\gamma_{xy}^2+(\sqrt{\gamma_x^2+\gamma_y^2}\pm M_I)^2+\Delta_d^2+\Delta_z^2\pm2\sqrt{(\epsilon_0\gamma_{xy}+\Delta_d\Delta_z)^2+(\sqrt{\gamma_x^2+\gamma_y^2}\pm M_I)^2(\epsilon_0^2+\Delta_z^2)}}.
\label{qp}
\end{equation}
In this case, we find that nodes disappear on all four bands once $|\gamma_y|+|M_I|<\Delta_0$. Furthermore, the presence of the interface allows for the interesting possibility that when $||\gamma_y|-|M_I||<\Delta_0<|\gamma_y|+|M_I|$ then nodes can be associated with only two bands, while the other two bands will not have nodes. We note that a $c$-axis oriented electric field can in principle be used to vary the magnitude of $\lambda_I$, allowing an opportunity to observe this effect. We also note that our
${\bf k}\cdot {\bf p}$-like theory  provides a hint as to why the observed Fermi surface shows no (or small) avoided crossing along the $k_x=0$ or $k_y=0$ directions. The observation that both the spin-orbit coupling and the interface potential will vanish at the ${\bf M}$-point suggests that these effects will be smaller than originally expected due to the proximity of the Fermi surface to the ${\bf M}$-point.

\section{BdG equations in the band basis}

To gain a deeper understanding on the origin of a nodeless $d$-wave gap, it is fruitful to examine the Bogoliubov deGennes (BdG) equations in the band basis as opposed to the orbital basis. Towards this end, we first note that the $8 \times 8$ BdG equations can be written in block diagonal form with two $4\times 4$ blocks. One of these blocks is
\begin{equation}
H_{BdG}=\left(
          \begin{array}{cccc}
            \epsilon_0+\gamma_{xy}&\gamma_x-i\gamma_y & 0& \Delta_d+\Delta_z \\ \gamma_x+i\gamma_y& \epsilon_0-\gamma_{xy}& -\Delta_d+\Delta_z& 0 \\ 0&-\Delta_d+\Delta_z&-\epsilon_0+\gamma_{xy}&\gamma_x+i\gamma_y \\ \Delta_d+\Delta_z &0&\gamma_x-i\gamma_y & -\epsilon_0-\gamma_{xy} \\
           
          \end{array}
        \right).
        \label{Hbdg}
        \end{equation}
The other block is found by taking $\Delta_i\rightarrow -\Delta_i$ and $\gamma_x+i\gamma_y\rightarrow \gamma_x-i\gamma_y$. Performing a unitary transformation that diagonalizes the normal part of the Hamiltonian yields 
\begin{equation}
H_{BdG}=\left(
          \begin{array}{cccc}
           \epsilon_0+\sqrt{\gamma_{xy}^2+\gamma_x^2+\gamma_y^2}&0 & -\frac{\Delta_z(\gamma_x-i\gamma_y)}{\sqrt{\gamma_{xy}^2+\gamma_x^2+\gamma_y^2}}& \Delta_d+\frac{\Delta_z \gamma_{xy}}{\sqrt{\gamma_{xy}^2+\gamma_x^2+\gamma_y^2}} \\  0&\epsilon_0-\sqrt{\gamma_{xy}^2+\gamma_x^2+\gamma_y^2}&  -\Delta_d+\frac{\Delta_z \gamma_{xy}}{\sqrt{\gamma_{xy}^2+\gamma_x^2+\gamma_y^2}}&\frac{\Delta_z(\gamma_x+i\gamma_y)}{\sqrt{\gamma_{xy}^2+\gamma_x^2+\gamma_y^2}}\\-\frac{\Delta_z(\gamma_x+i\gamma_y)}{\sqrt{\gamma_{xy}^2+\gamma_x^2+\gamma_y^2}}& -\Delta_d+\frac{\Delta_z \gamma_{xy}}{\sqrt{\gamma_{xy}^2+\gamma_x^2+\gamma_y^2}}&-\epsilon_0-\sqrt{\gamma_{xy}^2+\gamma_x^2+\gamma_y^2}&0 \\ \Delta_d+\frac{\Delta_z \gamma_{xy}}{\sqrt{\gamma_{xy}^2+\gamma_x^2+\gamma_y^2}}&\frac{\Delta_z(\gamma_x-i\gamma_y)}{\sqrt{\gamma_{xy}^2+\gamma_x^2+\gamma_y^2}}&0 & -\epsilon_0+\sqrt{\gamma_{xy}^2+\gamma_x^2+\gamma_y^2}\\
           
          \end{array}
        \right).
        \label{Hbdg}
        \end{equation}
In the band basis the BdG Hamiltonian has both intraband and interband pairing. The interband pairing requires a non-zero spin-orbit coupling to appear. Along the lines $k_x=0$ and $k_y=0$ (where nodes can appear)  the pairing is entirely interband. For weak splitting of the two Fermi surfaces (i.e. $v_{s0}k_F\lesssim\Delta_z$), the interband pairing can still gap out the low-energy states; for stronger spin-orbit coupling, however, the interband pairing potential is unable to overcome the band splitting, and so the single-particle gap has nodes along these directions. The presence of intraband pairing elsewhere in k-space confines the nodes to these lines.

\end{document}